# ENHANCING CYBERSECURITY DEFENSES: A MULTICRITERIA DECISION-MAKING APPROACH TO MITRE ATT&CK MITIGATION STRATEGY

Ihab Mohamed[1], Hesham A. Hefny[2], and Nagy R. Darwish[1]

[1]Department of Information Systems and Technology, Faculty of Graduate studies for Statistical Research (FGSSR), Cairo University, Egypt
[2]Department of Computer Science, Faculty of Graduate Studies for Statistical Research (FGSSR), Cairo University, Egypt

## ABSTRACT

*Cybersecurity is a big challenge as hackers are always trying to find new methods to attack and exploit system vulnerabilities. Cybersecurity threats and risks have increased in recent years, due to the increasing number of devices and networks connected. This has led to the development of new cyberattack patterns, such as ransomware, data breaches, and advanced persistent threats (APT). Consequently, defending such complicated attacks needs to stay up to date with the latest system vulnerabilities and weaknesses to set a proper cybersecurity defensestrategy. This paper aims to propose a defense strategy for the presented security threats by determining and prioritizing which security control to put in place based on combining the MITRE ATT&CK framework with multi-criteria decision-making (MCDM) techniques. This approach helps organizations achieve a more robust and resilient cybersecurity posture.*

## 1. INTRODUCTION

A cybersecurity risk assessment is an important tool that enables system stakeholders to assess the threats to their systems and take appropriate steps to mitigate those risks [1]. Cyber-attacks are becoming more common so most businesses need to be prepared to face them. To achieve this target, it is necessary to follow up security standards, to use and practice the best security tools for knowing what needs to be done to make systems more secure. However, such security plans are based on the knowledge of domain experts. It is hard to figure out how dangerous a vulnerability is in the context of security and risk management.

Avulnerability refers to defects or weaknesses in a system's design, operation, or implementation that could be used against the system to gain access to damage or to compromise sensitive data. Advanced persistent threats (APTs) are mainly complex attacks that are composed of several elementary attacks with no previous knowledge about what tactics they use. Inadequate cybersecurity strategies can lead to data breaches, identity theft, and other forms of cybercrime. Surveys show that organizations remain hesitant to implement effective cyber countermeasures despite the estimated cybercrime cost of $945 billion worldwide in 2020 [2].

According to an IBM study, 83% of organizations have had at least one data breach, and the average cost of a breach is $4.35 million in 2022, a 2.6% increase from 2021 [3]. The United

States ranks first in terms of the average cost of a data breach, with an average cost of $9.44 million, followed by the Middle East, with an average cost of $7.46 million, Canada, with an average cost of $5.64 million, and the United Kingdom, with an average cost of $5.05 million, and Germany, with an average cost of $4.85 million. Figure. 1 shows the average total cost of data breaches in million USD all over the world [3]. Also, according to the Allianz Risk Barometer, the top concerns for businesses globally in 2022 were: 44% for cyber incidents, 42% for business interruptions, and 25% for natural disasters. [4]. The significance of analyzing attack action correlations and forecasting hostile behavior, which enables proactive intrusion investigation and prevention, is an aspect that business solutions miss.

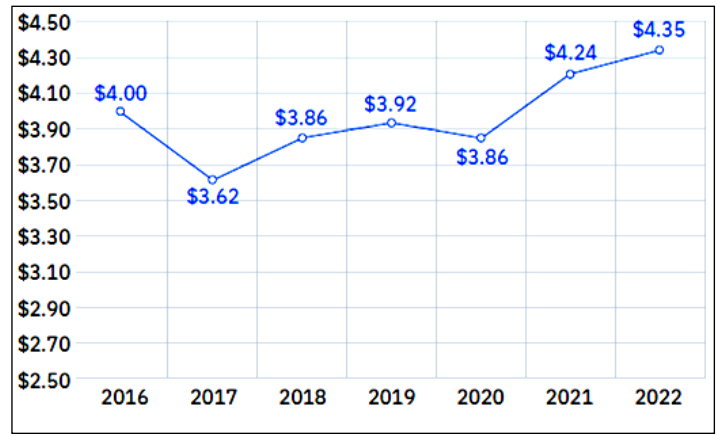

Figure1. The average total cost of data breach [3]

In an endeavor to mitigate this problem and overcome cyber threats and attacks, organizations are interested in developing plans to improve their security operations. This includes implementing security measures for monitoring, protecting, and applying policies and procedures to defend systems and data. To predict the strategies, techniques, and tactics of the adversary's behavior, the scientific community has been focusing on modeling cybersecurity attack patterns and methodologies based on reported incidents [5]. The Center of Threat Informed Defense (CID) builds on MITER ATT & CK® (which stands for Adversarial Tactics, Techniques, and Common Knowledge) and is considered an important threat-informed defense foundation used by security teams and vendors in their enterprise security operations. They have a map that shows how different controls and threat techniques work [6]. The adversary framework put forth by MITRE's ATT&CK framework is probably the most well-known defense strategy based on actual cyber-security threat observations.

The paper introduces an improved strategy to counter sophisticated cyber-security threats by formulating a prioritized response plan derived from the top ten recognized cybersecurity attack methods. This method is deemed highly efficient in leveraging common system vulnerabilities and weaknesses that are often exploited in significant cyber-attacks. The proposed approach allows combining Multi-Criteria Decision-Making Methods (MCDM) with MITRE ATT&CK framework to provide an ordered list of mitigation methods based on the priority ranking for each possible attack. This provides quick response and great flexibility for security management in establishing an effective defense at a low cost against any potential attack.

The rest of the paper is organized as follows: Section 2 basic concepts about vulnerabilities and cyberattack risks. Information security frameworks are presented in section 3. Section 4 presents the MITRE ATT&CK framework. MCDM methods are discussed in section 5. Related work found in the literature is given in section 6. Section 7 presents the proposed mitigation strategy based on the MCDM method. The formulation of the proposed MCDM method is given in section 8. An illustrative case study is demonstrated in section 9. Section 10 gives the conclusion and the future work.

## 2. VULNERABILITIES, THREATS, CYBERATTACKS

As per the National Institute of Standards and Technology (NIST), vulnerability is defined as a weakness in the computational logic (e.g., code) found in software and hardware components that, when exploited, results in a negative impact on confidentiality, integrity, or availability [7]. Threats are actions taken to gain a benefit from security breaches in a system and negatively affect it [8]. Cyberattacks are any kind of malicious activity that attempts to collect, disrupt, deny, degrade, or destroy information system resources or the information itself [9]. Risk is a function of threats exploiting vulnerabilities to impact operations and damage or destroy assets, see Figure. 2. A quantification of the risk factor is computed as follows [10]:

*Risk factor = (probability of threat) × (probability of vulnerability) × (impact of vulnerability)*
Therefore, any defense strategy should address the above four issues. Numerous sources offer information about quantifiable metrics and scoring measures on known vulnerabilities and exploitations. Such information can be used to guide defenders in targeting and prioritizing solutions. The following subsections give a brief overview of different vulnerability databases.

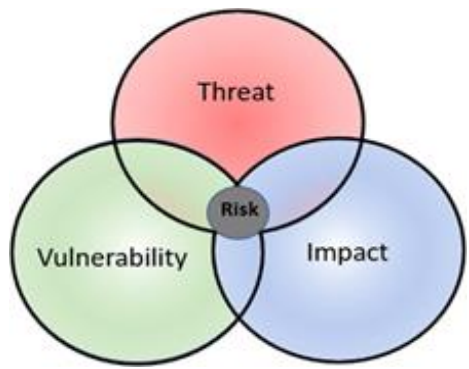

Figure 2. Risk factors [10]

### 2.1. National Vulnerability Database

National Vulnerability Database (NVD) is a U.S. government repository for standards-based vulnerability management. NVD was launched by the NIST in 2005 [11].

### 2.2. Common Vulnerabilities and Exposures

Common Vulnerabilities and Exposures (CVE) is a comprehensive cyber security vulnerability database that integrates all publicly available U.S. government vulnerability resources. It also provides references to industry resources [12]. The CVE list was launched by MITRE as a community effort in 1999.

### 2.3. Common Vulnerability Scoring System

Common Vulnerability Scoring System (CVSS) provides a means to "capture the principal characteristics of a vulnerability and produce a numerical score reflecting its severity" [13]. A quantification of the Exploitability metric is computed as follows [14]:

21

$$\text{Exploitability Metric} = 8.22 \times \text{Attack Vector} \times \text{Attack Complexity} \times \text{PrivilegesRequired} \times \text{User Interaction}.$$

### 2.4. Common Weakness Scoring System

Common Weakness Scoring System (CWSS) provided by MITRE is a mechanism for prioritizing software weaknesses that are present within software applications in a consistent and flexible manner [15][16]. The Centre for Threat-Informed Defence (CTID) has developed a methodology to use the adversary behaviors described in MITRE ATT&CK® to characterize the impact of vulnerabilities from CVE. This information can help defenders better assess the true risk posed by specific vulnerabilities in their environment [17].

## 3. INFORMATION SECURITY FRAMEWORKS

Security is defined as the absence of danger or threat [18]. To protect its operations, an organization has to maintain multiple layers of security in place, including physical security, personnel security, communications security, network security, and information security. The main pillars of information security are Confidentiality, Integrity, and Availability. Confidentiality ensures that authorized users only have access to the information they are authorized to [19]. Integrity ensures that the data is not modified or destroyed improperly and that it is available when needed [20]. Availability ensures that data is accessible and usable in a timely fashion. Information security standards provide the best practices and guidelines to ensure that data within the enterprise is protected from unauthorized access [19]. The following subsections present standards and frameworks of information security management that are essential for planning defense strategy.

### 3.1. ISO 27001 Framework

The International Organization for Standardization (ISO) launched the ISO 27001 standard as a model for setting up, running, and improving an information security management system (ISMS) by providing a framework for organizations to keep their data safe [20]. Adopting an ISO 27001 system is a strategic decision that affects the way a company operates and its relationships with other organizations. The ISMS system must balance the needs of the company with the security risks that must be managed.

### 3.2. NIST Cybersecurity Framework

According to NIST, Cybersecurity has multiple definitions [21]. The following are two common definitions: The first is "It is the process of protecting information by preventing, detecting, and responding to attacks". The second definition is: "It is the measures and controls that ensure confidentiality, integrity, and availability of the information processed and stored by a computer". The NIST Cybersecurity Framework is based on existing standards, guidelines, and practices and consists of three main components: the Framework Core, Implementation Tiers, and Profile [22].

### 3.3. Critical Security Controls (CIS)

CIS, by Sans Institute, provides a set of prioritized safeguards designed to lessen the most frequent cyber-attacks on systems and networks. These safeguards are based on legal, governmental, and policy frameworks, and are mapped to them so that they can be used as references. The update to CIS Controls v8 was prompted by the transition to cloud computing,

International Journal of Computer Networks & Communications (IJCNC) Vol.16, No.4, July 2024

virtualization, mobility, outsourcing, work-from-home policies, and evolving attacker strategies. This improved enterprise security by utilizing both full cloud and hybrid environments [23].

## 4. MITRE ATT&CK FRAMEWORK MODELLING APPROACH

The ATT&CK framework was developed by MITRE in 2013. It has a comprehensive, accessible database of attacks, techniques, and mitigation measures for enterprise, mobile, and industrial control systems (ICS) [5]. A structured taxonomy is provided by ATT&CK to define a variety of adversary activities. Formally, it is divided into three technical domains [23]:

- **Enterprise,** which describes behaviors on standard IT systems (e.g., Linux, Windows).
- **Mobile,** which focuses on mobile devices (e.g., Android, iOS).
- **ICS,** which relate more generally to cyber-physical systems.

Figure 3 is a diagram that outlines the MITRE ATT&CK relationships between an intrusion set, attack patterns, techniques, mitigations, and courses of action. The terms are arranged in a hierarchy, with intrusion sets at the top and courses of action at the bottom. Arrows depict various relationships between these terms. Overall, the MITRE ATT&CK framework provides a structured way to understand attacker behavior and helps defenders develop appropriate mitigation and response strategies [24]. The tactic describes the adversaries' overall goal, while the technique describes how they will achieve the goals. The procedure details the specific steps that need to be taken to carry out the technique [25]. By understanding the tactic, technique, and procedure (TTPs) of an attack, one can better predict how it will be performed and how to defend against it accordingly. The tactic and method abstraction in the model gives a way to identify adversary actions using both defensive and offensive cyber security [26]. To support relevancy and maximize impact on defenders, MITER ATT & regularly develops the framework by adding major new structures, features, and technologies[27]. The MITRE enterprise release in 2023 of ATT&CK for Enterprise (version 13) contains 14 tactics, 196 techniques, and 411 sub-techniques, [28][29].

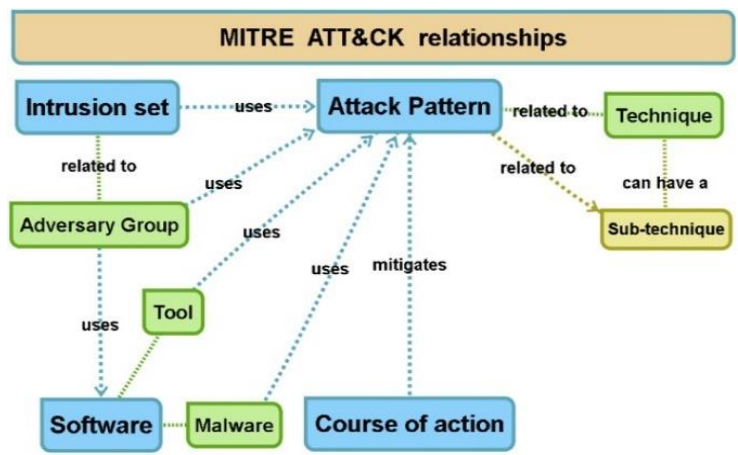

Figure 3. MITRE ATT&CK relationships [24]

## 5. MULTI-CRITERIA DECISION MAKING

Multi-criteria decision-making (MCDM) is one of the main decision-making tools that aims to decide the best alternative by considering more than one criterion in the selection process [29]. The weighting of criteria is a crucial component of multi-criteria decision analysis. There are

23

several methods for determining weights, which can be done manually based on the preferences and judgment of the decision-makers or automatically using developed methods and functions [30]. MCDM methods can be used to prioritize vulnerabilities or defense strategies based on their effectiveness against the TTPs listed in the ATT&CK matrix. Some examples of MCDM methods that have been used with MITRE ATT&CK in cybersecurity include [30], [31]:

- **Technique for Order Preference by Similarity to Ideal Solution (TOPSIS):**
  TOPSIS has been used to rank security controls based on their ability to mitigate specific TTPs in the ATT&CK matrix.
- **Analytic Hierarchy Process (AHP):**
  AHP has been used to prioritize defense strategies against the TTPs listed in the ATT&CK matrix based on their effectiveness, cost, and ease of implementation.
- **Preference Ranking Organization Method for Enrichment Evaluation (PROMETHEE):**
  PROMETHEE has been used to rank software vulnerabilities based on their criticality and potential impact on the organization's systems and processes.
- **Multi-Objective Optimization based on Ration Analysis (MOORA):**
  MOORA has been used to rank software vulnerabilities based on their criticality and the potential damage they could cause if exploited by an attacker.
- **Other very common MCDM methods are:**
  Weighted Sum Model (WSM) and Weighted Product Model (WPM) [32].

Adopting MCDM methods can help security teams prioritize their efforts and resources to address the most critical vulnerabilities or defense strategies based on the TTPs listed in the MITRE ATT&CK matrix.

## 6. RELATED WORK

As cybersecurity threats continue to evolve, researchers have been developing various methods to help mitigate these threats. V. Kumar et al., (2023) explored ways to prioritize software vulnerabilities and found that code execution vulnerabilities are more severe than authentication errors. This framework is then used to assess the severity of a vulnerability in a real-world case [33].

Mashima D. (2023) showed that recent attacks against critical infrastructure have been increasing in severity and stealth, and attackers may be hiding in the system after penetrating it to collect system information. To combat this, in-network deception technology is being used to create "apparent" but dummy devices that are deployed throughout the infrastructure to capture the attackers' reconnaissance activities. The effectiveness of this technology is studied based on the MITRE ATT&CK Matrix for industrial control systems. It has been found that it can effectively counter high-profile attacks that have recently been seen in the real world [34].

Heejung et al., (2022) illustrated how cyber threat intelligence (CTI) is the process of quickly collecting and analyzing data about potential security threats to find the nature and possible actions of a threat. A recent standard for doing this is the MITRE ATT&CK framework, but using this approach is time-consuming and often unsuccessful. The authors have developed a new way for classifying threats using easy data argumentation (EDA), and the authors found that it improved performance by 60–80% compared to using the standard method [35].

Octavian et al., (2022) introduced a dataset of 1813 CVEs and proposed models to automatically link a CVE to one or more techniques that can be used by cyber-security professionals to help mitigate attacks. The MITRE ATT&CK Framework The models are based on the text description from the CVE metadata. The paper also performs a qualitative analysis of the CVE descriptions [36].

Wenjun et al. (2022) proposed a threat modeling language specifically designed to help organizations analyze and protect their enterprise systems against cyber-attacks. The language is based on the MITRE Enterprise ATT&CK Matrix, and it focuses on describing system assets, attack steps, defenses, and asset associations. Testing has shown that the language is effective in describing real-world cyberattacks [37].

Liu Yet et al.,(2022) illustrated how defense models against adversarial attacks have grown significantly, but the lack of practical evaluation methods has hindered progress.Evaluation can be defined as looking for a defense model's lower bound of robustness given a budget number of iterations and a test dataset. The authors proposed a parameter-free adaptive attack evaluation method that addresses efficiency and reliability in a test-time-training fashion. By observing those adversarial examples to select a specific defense model, some regularities in their starting points and evaluation are observed. Furthermore, to approach the lower bound of robustness under the budget number of iterations, the author proposed an online statistics-based discarding strategy that automatically identifies and abandons hard-to-attack images. Notably, the author won first place out of 1681 teams in the CVPR 2021 white-box adversarial attacks on defense model competitions with this method [38].

An empirical study with expert cyber risk assessment practitioners was conducted to evaluate the adversary cost framework's validity and utility. However, focus is often put on the latter two components of cyber risk, vulnerability, and impact, due to the assessor being able to gather data about them reliably. Furthermore, many cyber risk assessment outputs are delivered in a qualitative or semi-quantitative format, which is incongruous with the output of other business functions, particularly at the board level [39].

Pell R et al., (2021) discussed how the gap between early 5G network threat assessments and an adversarial TTPs knowledge base for future use in the MITRE ATT&CK threat modeling framework can be bridged. The author identified knowledge gaps in the existing framework for key 5G technology enablers such as SDN, NFV, and 5G-specific signalling protocols of the core network. The authors studied how the domain-specific technique works and emulated this mapping in a pre-emptive fashion to facilitate a rigorous cyber risk assessment to support intrusion detection, and to design defenses based on common APT TTPs [40].

Ahmed Met et al., (2021) found that MITRE ATT&CK provides information on the motivations, capabilities, interests, and tactics of threat actors, which can be used to inform cyber risk characterization and assessment. The authors used this information to determine the likelihood of success of an attack against a healthcare organization based on the techniques and procedures of a known threat actor [2].

Choi S el al., (2021) proposed an automatic method for generating attack sequences that are credible and reflective of known cyber-attacks. This method is based on the tactics and techniques of MITRE ATT&CK and is applied to a dataset of industrial control system security [41].

Fang Z el al., (2021) suggested that one way to tackle data breach incidents is to develop a statistical model that can be used to predict data breach incidents for individuals and enterprises. To evaluate this model, the authors applied it to a dataset of enterprise-level breach incidents. The results show that the model is effective in predicting these incidents. Data breaches are a major cybersecurity problem, and they pose a risk to individuals' privacy. However, there is still much that is not understood about data breach risk, which makes it difficult to model and predict it [42]. Rahman M. et al. (2022) described the research that was conducted to help organizations make informed choices about the security controls to use to resist cyberattacks. The authors found that only a limited number of security controls can mitigate the many diverse types of adversarial techniques used in cyberattacks [43].

Most of the recent work on developing cybersecurity methodologies that adopt MCDM methods is directed toward quantification risk assessment based on subjective estimates of cyberattacks and existing system vulnerabilities. The MCDM approach helps in solving the vulnerabilities prioritization problem to develop a prioritization mitigation plan according to the ranked vulnerabilities [44].

Bhol et al., (2020) proposed a proactive fail-safe attitude based on MCDM, to rank organizations in terms of their security qualification. They used the security metrics: Susceptibility, Protection mechanism, Risk measurement, and Encounters, as the criteria with specific weights. Then they make a ranking comparison between the two MCDM approaches: AHP and ELECTRE III [45].

Horta et al., (2022) were concerned with improving a computer system's security posture based on the results of Breach Attacks Simulation (BAS). Their method was found to be robust and capable of producing coherent recommendations with a prioritized action plan based on context analysis. The heat maps presented were useful tools for the decision maker to understand which points needed more attention as well as to assess the detection capacity of their environment. According to their research work, to check organization security, a specific cyberattack has to be fabricated through simulation, then the type of symptoms that affected the system was monitored, and then, the required defense plan was determined using the MCDM method based on the ordered list of top 10 attack techniques announced by MITRE ATT&CK V11 [46].

Alain and S. M. Rahman proposed a proactive method to handle cybersecurity due to recent major data breaches. While existing defenses like firewalls and antivirus are helpful, attackers develop new methods. Security professionals must adopt innovative tools and techniques to combat these evolving threats. Our goal is to make infrastructure impenetrable and prevent attacks from any source. Early detection of vulnerabilities is our best defense to minimize damage [47].

A.Subil,andS. Nairintroduces a new way to measure network security, accounting for the fact that vulnerabilities change over time. Existing metrics don't fully address this dynamic nature, leading to poor risk management decisions. The proposed model considers factors like exploit and patch availability, which can significantly impact security depending on how vulnerabilities are connected and exploited. By incorporating the vulnerability age, the model provides a more realistic view of network security changes. It utilizes the Vulnerability Lifecycle model and CVSS metric to analyze how exploitability and impact evolve [48].

So far, there has not been a global catalogue of mitigation delivery techniques that includes a specific structure and focuses on a specific type of mitigation strategy used by organizations against threat actors. This lack of a specific mitigation strategy leaves the possibility of undetected attacks against an organization's infrastructure, as well as the identification of monitoring gaps[].

## 7. THE PROPOSED APPROACH

Based on the above discussion, there are two methods for handling the cybersecurity complex attack. The first is a *proactive* method that aims to enhance the defensive immunity of the organization using an appropriate cybersecurity framework. This method aims to help organizations improve their cybersecurity strategy by providing a documented set of policies and procedures that can help reduce the unknown vulnerabilities and misconfigurations that exist within an enterprise network. This method of cybersecurity takes a risk management perspective, helping to identify and manage all the risks associated with the network and systems. The second is the *reactive* method, which is mainly concerned with the defensive procedure that has to be followed in the event of an actual security attack.

Regarding the second method, the traditional procedure followed in realistic situations is that one determines the nature of the attack, determines its causes, and handles each attack pattern separately with its associated counter-mitigation techniques and control. This method is inefficient as it may apply the same mitigation techniques many times, causing overhead in time, effort, and cost.

This paper aims to propose an effective procedure for applying the second method. The following subsections present the main idea of the proposed approach and how to implement it as an MCDM problem.

### 7.1. Main Idea of the Proposed Approach

The proposed approach depends on the idea of taking advantage of knowing the relative importance of different security attacks. The importance score of each attack technique is calculated as the relative frequency of appearance of such attack technique in various categories of cyber threats.Table 1 [49], shows the top 10 techniques of attacks ordered by relative importance score.

Table 1. Top 10 ATT&CK techniques score V13[49]

| No. | Top Technique Score | TID | Name |
|---|---|---|---|
| 1 | 2.951542857 | T1053 | Scheduled Task/Job |
| 2 | 2.914285714 | T1059 | Command and Scripting Interpreter |
| 3 | 2.519447619 | T1562 | Impair Defenses |
| 4 | 2.330395238 | T1055 | Process Injection |
| 5 | 2.260190476 | T1036 | Masquerading |
| 6 | 2.253380952 | T1218 | Signed Binary Proxy Execution |
| 7 | 2.18777619 | T1574 | Hijack Execution Flow |
| 8 | 2.183333333 | T1047 | Windows Management Instrumentation |
| 9 | 2.116466667 | T1543 | Create or Modify System Process |
| 10 | 1.922504762 | T1112 | Modify Registry |

Table 1 shows the following information[50]:

International Journal of Computer Networks & Communications (IJCNC) Vol.16, No.4, July 2024

- **Rank**: This column shows the position (from 1 to 10) of each technique based on the percentage of malware samples exhibiting the behavior.
- **Score**: This column shows the percentage of malware samples analyzed in 2023 that exhibited the corresponding technique.
- **Technique ID**: This column displays the unique identifier for each MITRE ATT&CK technique, typically in the format "T1XXX".
- **Name**: This column provides a descriptive name for each technique.

Table 2 shows the list of the top 10 MITRE ATT&CK techniques observed in malware samples during 2023, ranked by their prevalence. Each of the attack techniques has in turn its own of security mitigations controls with equal importance within each attack technique [50].

Table 2. MITRE attack and mitigation techniques V13

| T1053 | T1059 | T1562 | T1055 | T1036 | T1218 | T1574 | T1047 | T1543 | T1112 |
|---|---|---|---|---|---|---|---|---|---|
| M1047 | M1049 | M1038 | M1040 | M1045 | M1042 | M1013 | M1040 | M1047 | M1024 |
| M1028 | M1040 | M1022 | M1026 | M1038 | M1038 | M1047 | M1038 | M1040 | |
| M1026 | M1045 | M1024 | | M1022 | M1050 | M1040 | M1026 | M1045 | |
| M1018 | M1042 | M1018 | | | M1026 | M1038 | M1018 | M1033 | |
| | M1038 | | | | | M1022 | | M1028 | |
| | M1026 | | | | | M1044 | | M1022 | |
| | M1021 | | | | | M1024 | | M1018 | |
| | | | | | | M1051 | | | |
| | | | | | | M1052 | | | |
| | | | | | | M1018 | | | |

In Table 2, it is interesting to note that the same security mitigation controls can be adopted with various techniques of attacks. Hence, the novelty of the idea of our proposed approach lies in the fact that it aims to follow an effective procedure to quickly repel the security attack by applying a limited number of effective counter-mitigation steps. This helps to avoid wasting time in analyzing the type of complex cyber-attack and the different categories of threats that compose it. We have a solid fact about the top $n$ possible techniques of attacks that are quite likely to be adopted in any complex cyber-attack with arbitrary multiple combinations of categories of threats. Accordingly, the defense strategy for an arbitrary complex cyber-attack should depend on adopting security mitigation controls based on their prioritization over all the top 10 attack techniques. Thus, this is a direct MCDM problem where the top 10 attack techniques act as criteria and the security mitigation controls act as alternatives.

Therefore, the proposed approach in this paper follows the second method of handling the cybersecurity complex attack by combining MCDM methods with the MITRE Attack cybersecurity framework version 13.0 announced in 2023 with its recent list of top techniques of attacks. The list includes different techniques of attacks that are found to be adopted in each category of cyber threat, e.g. Ransom category or Data Breach category, etc. Also, for each attack, the list of counter-security mitigation controls to be used against it is also declared [51], [49]. It is interesting to mention that the same techniques of attacks are found to be adopted in various categories of cyber threats.

### 7.2. MCDM Problem Formulation

The following are the steps for formulating our MCDM problem:

**Step 1:**

The first step is to construct the decision matrix. The columns and rows of the matrix correspond to the criteria and the alternatives respectively. In our case, the top 10 attack techniques are the criteria, while the security mitigation controls are the alternatives. The decision matrix $MT_{m \times n}$ is given in (1).

$$MT = \begin{array}{c} \\ M_1 \\ M_2 \\ \\ M_i \\ \\ M_m \end{array} \begin{array}{cccc} T_1 & T_2 & ..T_j... & T_n \\ \left[ \begin{array}{cccc} mt_{11} & mt_{12} & & ...mt_{1n} \\ mt_{21} & mt_{22} & & ...mt_{2n} \\ & & ... & \\ & & ...... mt_{ij} & \\ mt_{m1} & mt_{m2} & & ...mt_{mn} \end{array} \right] \end{array} \quad (1)$$

Where $mt_{ij}$ is the performance score of the ith security mitigation control with respect to the jth attack technique, $m$ is the number of mitigation controls and $n$ is the number of techniques of attacks.

**Step 2:**

The second step is to assign a weight $w_j$ to each attack technique $T_j$ to reflect its importance among other techniques.

**Step 3:**

The third step is to compute the total score for each mitigation control $M_i$ overall attacks $T_j$ as given in (2).

$$M_i^{wsm} = \sum_{j=1}^{n} mt_{ij} \ w_j \quad , for \ i = 1,2,...,m \quad (2)$$

**Step 4:**

The fourth step is to prioritize the security mitigation controls by ranking the vector $M_i^{wsm}$ from the highest performance score to the lowest one.

Thus, the proposed defense plan against any arbitrary complex attack is to apply the ranked security mitigation controls sequentially. It is likely expected to abort the complex cyber-attack after applying a few of the top prioritized mitigation controls.

A demonstration of the applicability of the proposed approach is given through a case study. Assuming an organization has been subjected to a ransomware cyberattack. To apply the proposed MCDM approach we must select appropriate values for the *n* columns and *m* rows in (1). In the context of the Enterprise-attack mitigation v13, the decision matrix involves a combination of $n=1,2,...$up to 1238 attack techniques/sub techniques and $m=1,2,...$up to 43 mitigation techniques. This implies that there are 1238 criteria ($n$) and 43 alternatives ($m$). From a practical point of view, it is adequate to deal with the top 10 attack techniques, i.e. $n=10$, and the top 18 security mitigation controls, i.e. $m=18$. The weight $w_i$ of each attack technique $T_j$ is given in Table 1. Consider that all mitigation controls that belong to a specific attack technique have equal priority of application. Therefore, we will assign the same value of performance score $mt_{ij} i=1,2,...,m$ as follows:

$$mt_{ij} = \frac{w_j}{|mt_i|} \quad (3)$$

Where, $|mt_i|$ is the number of mitigation controls that belong to attack technique $T_j$ as given in Table 3. Thus, $|mt_1|=4$, $|mt_2| = 7$, ... $|mt_9| = 7$, $|mt_{10}|=1$.

Applying (2) we get the overall decision matrix as given in Table 3. It is clear that the last column is sorted based on the highest score of the considered mitigation controls.

Table 3. The Final Decision Matrix of Mitigation Techniques Score

| source ID | 2.951543 | 2.914286 | 2.519448 | 2.330395 | 2.26019 | 2.253381 | 2.187776 | 2.183333 | 2.116467 | 1.922505 | |
|---|---|---|---|---|---|---|---|---|---|---|---|
| Mitigation | T1053 | T1059 | T1562 | T1055 | T1036 | T1218 | T1574 | T1047 | T1543 | T1112 | Total Mitigations |
| M1026 | 0.590309 | 0.416327 | 0 | 1.165198 | 0 | 0.563345 | 0 | 0.545833 | 0 | 0 | 3.281011 |
| M1038 | 0 | 0.416327 | 0.629862 | 0 | 0.753397 | 0.563345 | 0.218778 | 0.545833 | 0 | 0 | 3.127541 |
| M1024 | 0 | 0 | 0.629862 | 0 | 0 | 0 | 0 | 0 | 0 | 1.922505 | 2.552367 |
| M1040 | 0 | 0.416327 | 0 | 1.165198 | 0 | 0 | 0.218778 | 0.545833 | 0.302352 | 0 | 2.648487 |
| M1018 | 0.590309 | 0 | 0.629862 | 0 | 0 | 0 | 0.218778 | 0.545833 | 0.302352 | 0 | 2.287134 |
| M1022 | 0 | 0 | 0.629862 | 0 | 0.753397 | 0 | 0.218778 | 0 | 0.302352 | 0 | 1.904389 |
| M1045 | 0 | 0.416327 | 0 | 0 | 0.753397 | 0 | 0 | 0 | 0.302352 | 0 | 1.472076 |
| M1047 | 0.590309 | 0 | 0 | 0 | 0 | 0 | 0.218778 | 0 | 0.302352 | 0 | 1.111439 |
| M1042 | 0 | 0.416327 | 0 | 0 | 0 | 0.563345 | 0 | 0 | 0 | 0 | 0.979672 |
| M1028 | 0.590309 | 0 | 0 | 0 | 0 | 0 | 0 | 0 | 0.302352 | 0 | 0.892661 |
| M1050 | 0 | 0 | 0 | 0 | 0 | 0.563345 | 0 | 0 | 0 | 0 | 0.563345 |
| M1049 | 0 | 0.416327 | 0 | 0 | 0 | 0 | 0 | 0 | 0 | 0 | 0.416327 |
| M1021 | 0 | 0.416327 | 0 | 0 | 0 | 0 | 0 | 0 | 0 | 0 | 0.416327 |
| M1033 | 0 | 0 | 0 | 0 | 0 | 0 | 0 | 0 | 0.302352 | 0 | 0.302352 |
| M1013 | 0 | 0 | 0 | 0 | 0 | 0 | 0.218778 | 0 | 0 | 0 | 0.218778 |
| M1044 | 0 | 0 | 0 | 0 | 0 | 0 | 0.218778 | 0 | 0 | 0 | 0.218778 |
| M1051 | 0 | 0 | 0 | 0 | 0 | 0 | 0.218778 | 0 | 0 | 0 | 0.218778 |
| M1052 | 0 | 0 | 0 | 0 | 0 | 0 | 0.218778 | 0 | 0 | 0 | 0.218778 |

Now, considering such a complex cyber-attack, the organization needs to apply the mitigation controls according to their calculated rank sequentially. It is likely expected to abort the complex attack before completing the whole 18 mitigation controls. However, again, values of $n$ and $m$ may be set to larger values if needed under certain circumstances.

## 8. ILLUSTRATIVE EXPERIMENT

This section presents a simulation experiment to evaluate the effectiveness of cybersecurity defences of the proposed approach against malicious behaviors. The experiment has been conducted using Caldera TM simulator[52]. Caldera is a framework from MITRE that allows for automated adversary emulation. The components of the experiment are as follows:

- **Caldera Server**: This is the central core of the experiment. It acts as the emulation platform, coordinating the simulated attacks.
- **Windows Virtual Machines (Labelled Devices)**: The three VM Machines connect to devices labeled "DC" (domain controller) and "Client Win 10" (client machine running Windows 10). These represent the target systems under attack in the simulation
- **Agents**: These agents are installed on the target machines. They relay information between Caldera and the target machines, allowing Caldera to execute commands and monitor system behavior. There are three "Agents" indicating three target machines.

Figure 4 depicts the components of the experiment.

The experiment utilized data from a Breach Attack Simulator (BAS) campaign. This suggests the Caldera platform may be leveraging pre-defined attack scenarios or behaviors during the emulation. MITRE ATT&CK is a knowledge base of adversary tactics and techniques, and the experiment specifically focused on the top 10 techniques of malicious behaviors. This indicates the attackers in the simulation will employ tactics that are commonly used in real-world attacks.

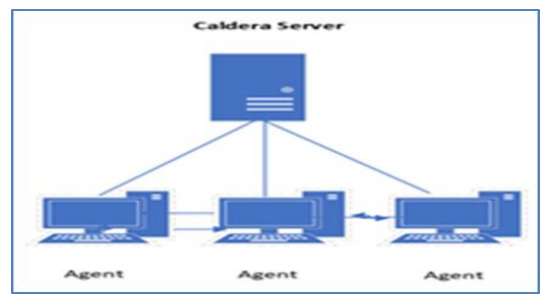

Figure 4. Caldera experiment

Thus, we employ the Caldera2 platform to simulate a virtual complex cybersecurity attack campaign that is composed of top10 attacktechniques. The experiment simulates a two-pronged cybersecurity attack. The first attack is "PowerShell Cmdlet Scheduled Task" which enables adversaries that may abuse the "Windows Task Scheduler" to perform task scheduling for the initial or recurring execution of malicious code. The second attack is "Visual Basic for Applications (VBA) and VBScript". VBA is an event-driven programming language built into Office applications. VBA enables documents to contain macros used to automate the execution of tasks and other functionalities on the host.

After the simulation of each attack technique, the adopted platform recorded observations in the logs to determine whether each attack was discovered, blocked, or not. Table 4 shows the infection status of the system after applying the simulated cyberattack. It is clear from Table 4 that the virtually generated cyberattack is composed of the top 2 attacks reported in Table 3, i.e T1053 & T1059. The authors manually consolidated all the simulated information resulting in an input spreadsheet. The authors then built an action plan with control items according to preferences that make sense for the tested environment. The control items were correlated based on the top 10 techniques from the MITRE ATT&CK framework found in Table 3. Table 5 shows the result of the simulation experiment after applying the proposed prioritized mitigation techniques. The results ensure the effectiveness of the proposed method to block the mentioned

International Journal of Computer Networks & Communications (IJCNC) Vol.16, No.4, July 2024

cyberattack.Also, Table 6 shows the advantages of using MCDM methodscompared with the traditional manual one.

Table 4. The Experimental Result before applying the proposed approach.

| Enterprise | Platform | Technique_Id | Technique Name | Protection Status | Top MITRE ATT&CK Techniques |
|---|---|---|---|---|---|
| Test Lab | Windows | T1053.005 | Powershell Cmdlet Scheduled Task | Not blocked | Experimental-1 |
| Test Lab | Windows | T1059.005 | Extract Memory via VBA | Not blocked | Experimental 1 |

Table 5. The Experimental Result after applying the proposed approach.

| Enterprise | Platform | Technique_Id | Technique Name | Protection Status | Top MITRE ATT&CK Techniques |
|---|---|---|---|---|---|
| Test Lab | Windows | T1053.005 | Powershell Cmdlet Scheduled Task | blocked | Experimental-2 |
| Test Lab | Windows | T1059.005 | Extract Memory via VBA | blocked | Experimental-2 |

Table 6. The manually/randomly – MCDM Comparison

| Factor | Traditional Manual/Random method | The proposed method |
|---|---|---|
| Approach | Mitigation techniques are applied manually or randomly, without considering any criteria or priorities. | Mitigation techniques are selected and applied based on MCDM method that prioritizes mitigation techniques of the top $n$ cyberattack |
| Effectiveness | Mitigation techniques may not be applied effectively or efficiently and may not be aligned with the organization's risk tolerance. | Mitigation techniques are applied effectively and efficiently and are more likely to be aligned with the organization's risk tolerance. |
| Transparency | It can be difficult to track and measure the effectiveness of mitigation techniques that are applied manually or randomly. | It is easier to track and measure the effectiveness of mitigation techniques that are applied using the MCDM WSM, as the criteria and weights are clearly defined. |
| Complexity | Manual/random implementation of mitigation techniques is less complex, but it can be high expensive and less effective. | MCDM WSM is done once for the selected top $n$ cyberattack techniques reported in literature. Then, it is applied in quite simple and systematic way for arbitrary types of complex cyberattacks with high level of effectiveness and likelihood to block the attack using few number of mitigation techniques. |

## 9. CONCLUSION

This paper demonstrates an approach for developing a prioritized mitigation plan to assess powerful cyber security. The proposed approach adopts an effective defense strategy based on MCDM WSM for ranking the security mitigation techniques of the recent top complex cyber-attacks declared by MITRE Attack Cyber security Framework V13.0. Once ranking, then a priority list of security mitigation controls is obtained based on their effectiveness and coverage through the top listed cyber-attacks.The security mitigations are applied sequentially according to their priorities.It is quite likely that after the application of a few mitigations of the ranked list, the cyber threat is likely to be eliminated. This proposed defense strategy allows taking quick action to fix security vulnerabilities and reduce time, overhead, and effort.We believe that this strategy is quite effective and least expensive in time and effort against different types of attacks. In future work, we plan to apply the proposed approach using other MCDM methods.

## CONFLICT OF INTEREST

The authors declare no conflict of interest.

International Journal of Computer Networks & Communications (IJCNC) Vol.16, No.4, July 2024[36]   O. Grigorescu, A. Nica, M. Dascalu, and R. Rughinis,(2022)  "CVE2ATT&CK: BERT-Based Mapping of CVEs to MITRE ATT&CK Techniques," Algorithms, vol. 15, no. 9, Sep. 2022, doi: 10.3390/a15090314.

[37]   W. Xiong, E. Legrand, O. Åberg, and R. Lagerström, (2022) "Cyber security threat modeling based on the MITRE Enterprise ATT&CK Matrix," SoftwSyst Model, vol. 21, no. 1, pp. 157–177, doi: 10.1007/s10270-021-00898-7.

[38]   Y. Liu, Y. Cheng, L. Gao, X. Liu, Q. Zhang, and J. Song, (2022)"Practical Evaluation of Adversarial Robustness via Adaptive Auto Attack." [Online]. Available: https://github.com/liuye6666/adaptive auto attack

[39]   Derbyshire, R., (2022)" Anticipating Adversary Cost: Bridging the Threat-Vulnerability Gap in Cyber Risk Assessment", Doctoral dissertation, Lancaster University.

[40]   R. Pell, S. Moschoyiannis, E. Panaousis, and R. Heartfield, (2021)"Towards Dynamic Threat Modelling in 5G Core Networks Based on MITRE ATT&CK,". https://doi.org/10.48550/arXiv.2108.11206

[41]   S. Choi, J.-H. Yun, and B.-G. Min,(2021)"Probabilistic Attack Sequence Generation and Execution Based on MITRE ATT& CK for ICS Datasets," in Cyber Security Experimentation and Test Workshop, Aug. , pp. 41–4.,doi:10.1145/3474718.3474722

[42]   Z. Fang, M. Xu, S. Xu, and T. Hu, (2021)"A Framework for Predicting Data Breach Risk: Leveraging Dependence to Cope with Sparsity," IEEE Transactions on Information Forensics and Security, vol. 16, pp. 2186–2201, doi: 10.1109/TIFS.2021.3051804.

[43]   M. R. Rahman and L. Williams,(2022)"An investigation of security controls and MITRE ATT\&CK techniques," .

[44]   Gourisetti S. N. G, M. Mylrea and H. Patangia, (2022) "Cybersecurity vulnerability mitigation framework through empirical paradigm: Enhanced prioritized gap analysis", Future Generation Computer Systems, vol. 105, pp. 410-431.

[45]   S. G. Bhol, J. R. Mohanty, and P. K. Pattnaik,(2020) "Cyber Security Metrics Evaluation Using Multi-criteria Decision-Making Approach," in Smart Innovation, Systems and Technologies,  vol. 160, pp. 665–675.  doi: 10.1007/978-981-32-9690-9_71.

[46]   A. Horta, R. Holanda, and R. Marinho,(2022) "A Multi-criteria Approach to Improve the Cyber Security Visibility Through Breach Attack Simulations.", Proceedings of the 22nd     Brazilian Symposium on Information and Computational Systems Security, pp. 330-343, doi: https://doi.org/10.5753/sbseg.2022.224454

[47]   Alain Loukaka1 and S. M. Rahman,(2017)" Discovering New Cyber Protection Approaches From A Security Professional Prospective", vol. 9, pp. 13–25.  doi: 0.5121/ijcnc.2017.9402

[48]   Abraham, Subil, and Suku Nair.(2015) "A predictive framework for cyber security analytics using attack graphs." International Journal of Computer Networks & Communications (IJCNC)ISSN:0974-9322; 0975-2293. "Calculator", Access (2023.Feb)" Available at: https://github.com/center-for-threat-informed-defense/top-attack techniques/

[49]   "Accessing ATT&CK Data," (2023). https://attack.mitre.org/resources/working-with-attack/

[50]   D. J. Bodeau, R. D. Graubart, R. M. Mcquaid, and J. Woodill, (2018) "Cyber Resiliency Metrics, Measures of Effectiveness, and Scoring,", MITER Technical Report, Public Release Case number 18-2579.

[51]   "MITRE Caldera" ( 2023)., https://github.com/mitre/caldera/
35